\documentclass[aps,prl,twocolumn,groupedaddress]{revtex4}
\usepackage{amsmath,amssymb,graphicx}
\usepackage{multirow}
\begin{document}
\title{Hole Localization in Molecular Crystals From Hybrid Density Functional Theory}

\author{Na Sai}
\email[]{nsai@physics.utexas.edu}
\author{Paul F. Barbara}
\email[]{Deceased}
\affiliation{Center for Nano and Molecular Science and Technology, The University of Texas at Austin, Austin, Texas, 78712, USA}
\author{Kevin Leung}
\email[]{kleung@sandia.gov}
\affiliation{Surface and Interface Sciences Department, MS1415, Sandia National Laboratory, Albuquerque, New Mexico 87185, USA} 
\pacs{71.10.-w, 71.15.Mb, 71.38.Ht,31.15.-p}
\date{\today}
\begin{abstract}
We use first-principles computational methods to examine hole trapping in organic molecular crystals.  We present a computational scheme based on  the tuning of the fraction of exact exchange in hybrid density functional theory to eliminate the many-electron self-interaction error.  With small organic molecules, we show that this scheme gives accurate descriptions of ionization and dimer dissociation. We demonstrate that the excess hole in perfect molecular crystals form self-trapped molecular polarons. The predicted absolute ionization potentials of both localized and delocalized holes are consistent with experimental values.
\end{abstract}
\maketitle 

Charge localization in organic solids and interfaces has important implications for the functioning of organic devices~\cite{Silinsh, Pope}. It has been long suggested that charge carriers in organic molecular crystals may self-trap to form localized small polarons~\cite{Holstein, Emin, Silinsh, Pope} through interaction with the surrounding electron and lattice. Models based on the concept of polaron have led to theoretical understanding of charge transport of these materials (see e.g.~\cite{Silinsh, Kenkre}). On the other hand, direct evidence and atomic scale probe of localized charge states in organic crystals remain lacking. {\it Ab initio} computations have recently been summoned to provide accurate quantitative description of polaron state in covalent solids~\cite{Ramo, Franchini}. 
 In this work, we apply hybrid density functional theory (DFT) with carefully calibrated admixtures of exact exchange to molecular crystals and conclusively demonstrate from first principles the existence of {\it self-trapped hole polaron} in defect-free molecular crystals.  Our choice to work with small organic molecules permits systematic analysis of DFT exchange correlation (xc) functionals in both gas and solid phases. It also allows predictions of the absolute ionization potentials (IP) of both polaronic localized and free delocalized hole state in solids, revealing large errors in semilocal (sl) DFT functionals that to our knowledge have never been discussed in the literature. This successfully treatment of hole localization and delocalization properties will aid the understanding of charge trapping~\cite{Kaake, Bolinger} and carrier mobility~\cite{Gershenson} that are important for organic electronics.

A key prerequisite to such studies is the correct choice of DFT functionals. Previous theoretical studies reveal an extreme sensitivity of charge localization to theoretical
approximations~\cite{Grossman,Lany,Wu, Ramo, Franchini, Kimmel}.  Standard semilocal 
functionals for the xc energy, such as the local density
approximation (LDA) and generalized gradient approximation (GGA), have
been successful in predicting atomic structures and electronic properties
of many molecular complexes and solid state materials. But they tend
to fail in systems involving fractional charges, e.g. they dissociate molecular ion H$_2^+$ into 2 H$^{0.5+}$ rather
than into H$^+$ and H.  
The problem stems from self-interaction
error (SIE) in the standard density
functionals~\cite{Cohen}. For a one electron
system SIE is conveniently described as the incomplete cancellation
of spurious self-exchange and
self-Coulomb energy~\cite{Perdew81}. A more general many-electron SIE  (also
known as the delocalization error)~\cite{Ruzsinszky, Sanchez} arises from the incorrect behavior of the total energy as a functional of the average number of electrons. While the exact energy varies linearly between adjacent integers~\cite{Perdew82}, commonly used xc functionals give rise to convex or concave behavior~\cite{Perdew07}, which tends to overstabilize either the delocalized or the localized electron density respectively. This will
clearly impair accurate prediction of charge trapping and localization in organic solids. To achieve a many-electron-SIE-free situation, the ground state total energy $E(N)$ should be linear in the electron number $N$ 
and has a realistic slope~\cite{Vydrov}.  

In this letter, we achieve this condition by tuning the mixing parameter $\alpha$ in a hybrid functional, which combines a fraction of the nonlocal exact Hartree-Fock (HF) type
exchange with the semilocal exchange, 
\begin{equation}
E_{x} = \alpha E^{\rm HF}_x+ (1-\alpha) E^{\rm sl}_x  
\end{equation} 
 where $0<\alpha \le1$ defines the mixing coefficient. The widely used PBE0 hybrid functional~\cite{Perdew96, Adamo} sets $\alpha=0.25$ and leads to band gaps and atomic energies in better agreement with the
experiments. However, we will show that the global PBE0 ($\alpha=0.25$), owing to the residual SIE, gives an erroneous dissociation limit in the benzene dimer cation and a
delocalized charge in molecular solids with excess charge.   
By contrast, our hybrid functional leads to correct
dissociative behavior of molecular dimer cations and yields
{\it single molecular polaron}-like hole localization in molecular solids. We demonstrate the validity
of this approach for a number of small organic molecules with a focus on polyacenes. 

We have performed the calculations in the {\tt VASP} program using 
projector-augmented wave potentials~\cite{VASP,PAW}. All the calculations
are spin unrestricted and without symmetry constraint. 
We use a kinetic cutoff of 400 eV and
 $\Gamma$ k-point Brillouin zone sampling.  
For charged systems we add the monopole correction~\cite{Makov}. 
Because the molecular solids considered in this work exhibit high symmetry and obvious physical boundaries
(terminated as intact molecules), the large and unscreened quadruple
correction~\cite{Leung} can also be determined to yield absolute ionization potentials.

\begin{figure}
\includegraphics*[width=8.5cm]{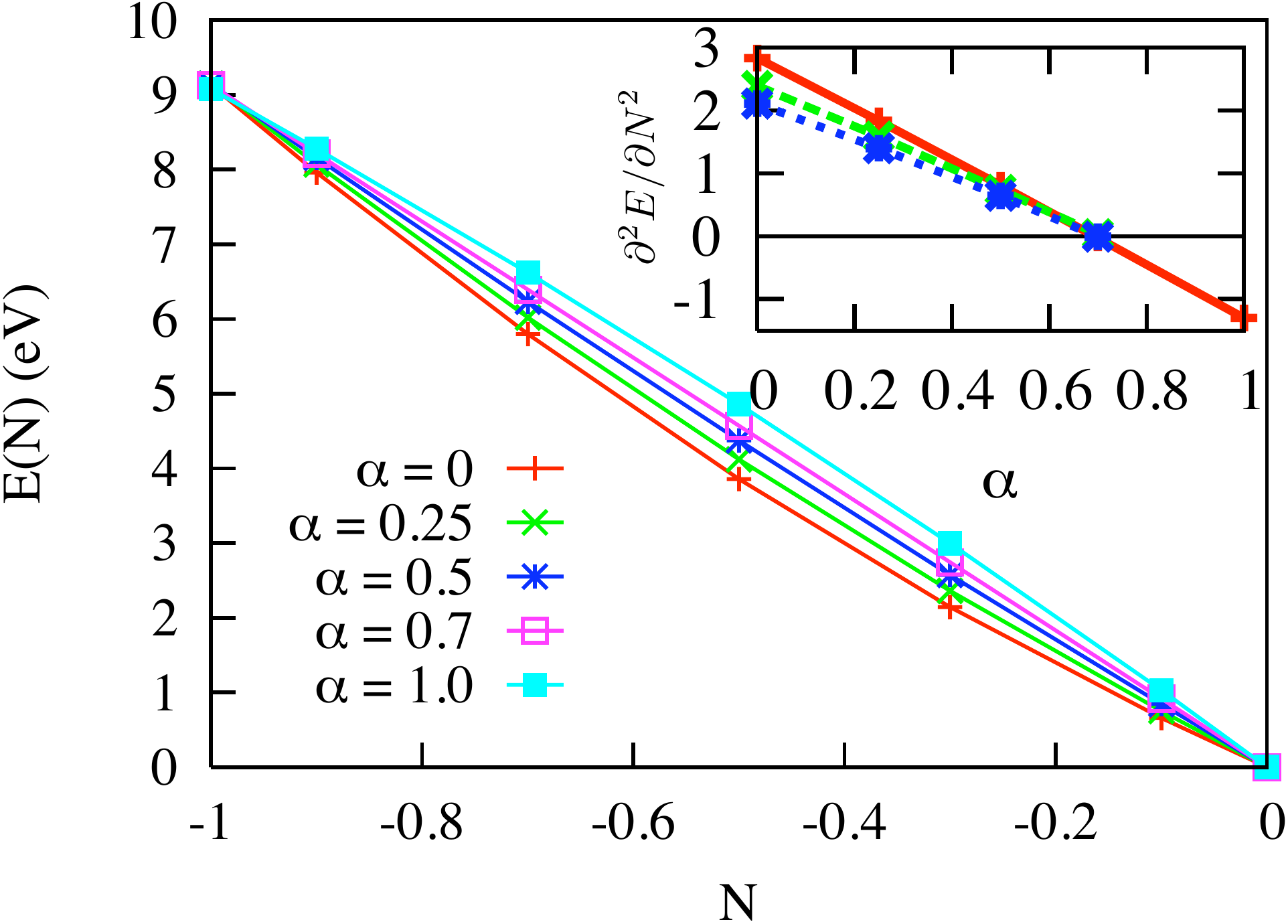}
\caption{ Total energy of the benzene molecule as a function of the
electron number $N$; the energy is zeroed out at $N=0$. $\alpha =0 $ and $\alpha=0.25$ correspond to the PBE and the global PBE0 hybrid  functional, 
and $\alpha = 1.0$ corresponds to the full HF exchange with PBE correlations. Inset: the curvature of the total energy with respect to the electron number as a function of $\alpha$ for benzene (solid red), naphthalene (dashed green), and anthracene(doted blue). } 
\label{ben_ex}
\end{figure} 
First we consider isolated polyacene molecules. 
Fig.~\ref{ben_ex} illustrates the total energy $E(N)$ of benzene as a function of the electron number $N$, where $N$ varies from $M-1$ (cation) to $M$ (neutral). The PBE curve ($\alpha = 0$) shows an incorrect convex behavior as the fraction of hole charge increases, and PBE0 ($\alpha = 0.25$) improves only slightly upon PBE but remains
convex. As $\alpha$ increases, the energy curve becomes increasingly more linear
while at $\alpha = 1.0$, i.e., with the full HF exchange (plus PBE correlation), it
becomes concave. The linear dependence of the curvature $\partial^2 E(N)/\partial N^2$ on $\alpha$ (Fig.~\ref{ben_ex} inset) enables us to determine the critical value $\alpha_c$ that fulfills the linear criterion. For all three polyacenes, the condition is satisfied practically at the same value $\alpha_c= 0.7$ (much higher than $\alpha=0.25$ used in PBE0) suggesting no direct dependence of $\alpha_c$ on the conjugation length. At $\alpha= 0.7$, the gas phase ionization energies $I_g= 9.13$, 8.01, and 7.23 eV for benzene, naphthalene, and anthracene, in agreement with the respective experimental values of 9.24, 8.14, and 7.45 eV~\cite{Handbook}. The HOMO-LUMO gaps of 9.2, 8.01 and 6.85 eV are also in agreement with the respective experimental values of 9.98, 8.0, and 6.76 eV~\cite{Handbook}. 

For a benzene dimer cation (C$_6$H$_6$)$^+_2$ in the dissociation limit, we have found that both PBE and PBE0 (even at $\alpha=0.65$) yield
a symmetric dissociated state with $+0.5e$ on each
molecule. By contrast, the hybrid functional at $\alpha=0.7$ yields an asymmetric dissociated state, i.e., a neutral (C$_6$H$_6$)$^0$ and
a cation (C$_6$H$_6$)$^+$, as expected.   

With the protocol established in molecules, we proceed with the calculation of hole state in solids. Using a $2\times2\times2$ extended supercell consisting of 384 atoms and starting out with a perfect crystal at equilibrium, we insert an excess hole into the system by removing an electron and allow the atomic positions to fully relax while keeping the lattice parameters fixed at
the experimental values~\cite{crystal}

\begin{figure}
\includegraphics*[width=7 cm]{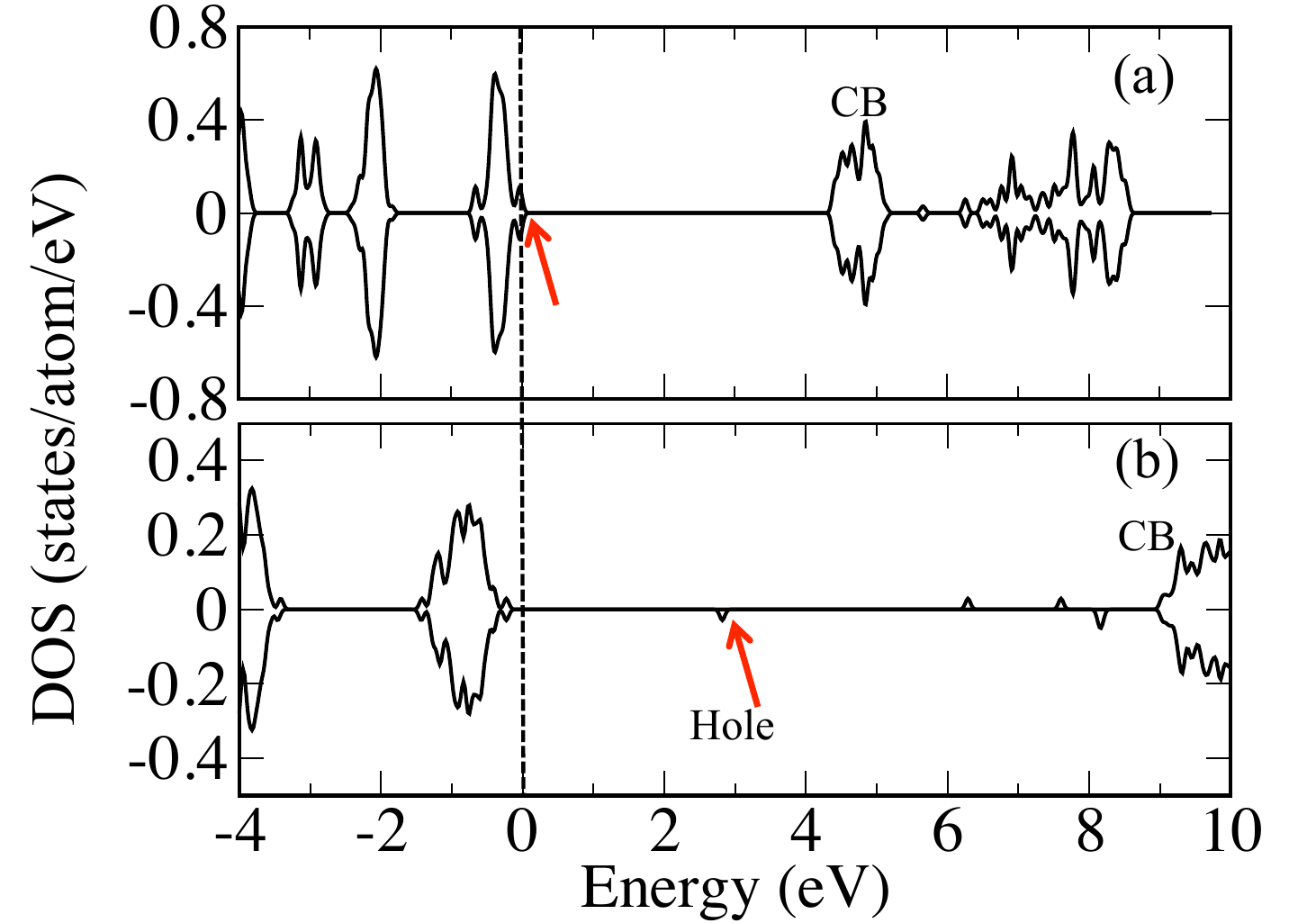}
\caption{Density of states of the benzene crystal $2\times2\times2$
supercell with an excess hole calculated from
(a) PBE and (b) the hybrid functional at $\alpha = 0.7$. The arrows guide the eyes
for the hole states. The molecular distortion related states are visible in the gap in (b).}
\label{dos}
\end{figure}
\begin{figure}
\includegraphics*[width=6 cm]{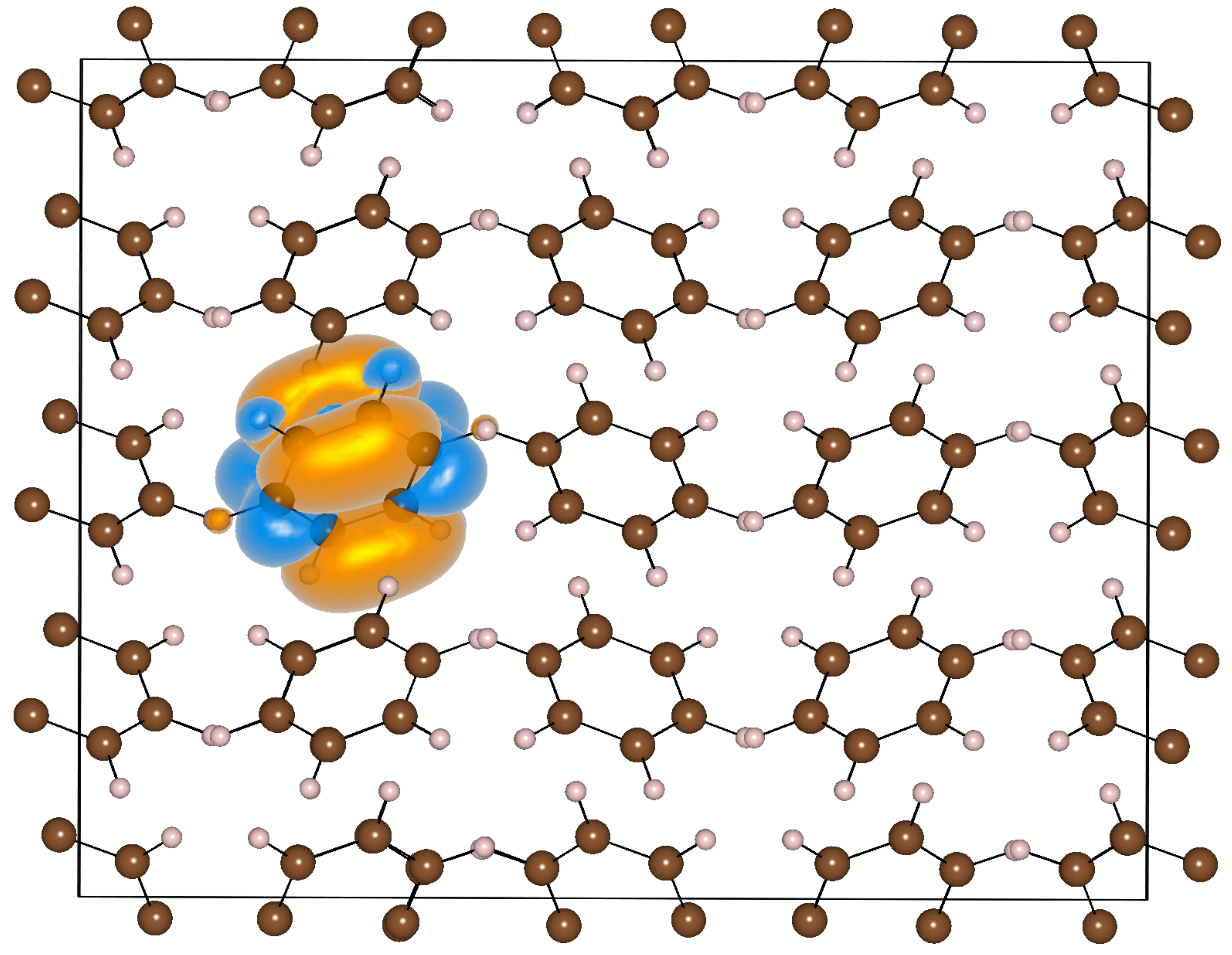}
\caption{ Spatial distribution of the excess hole state
in the benzene crystal calculated from the hybrid $\alpha = 0.7$ functional. The hole
is shown self-trapped around a single benzene site. }
\label{loc}
\end{figure}

 
The density of states (DOS) for the positively charged benzene crystal is shown in Fig~\ref{dos}. The PBE hole state splits into half occupations at the
top of the valence band leading to  degenerate spin up and spin down states as shown in Fig.~\ref{dos} (a). It is clear that this hole state is fully
delocalized over the entire lattice. In the case of PBE0 ($\alpha = 0.25$), although the hole state lies
well above the valence band, our calculation shows that it fully
delocalizes in the crystal regardless of the structural relaxations. In contrast, the hybrid $\alpha=0.7$ functional, which exhibits no SIE for an isolated C$_6$H$_6$, predicts that the hole state occupies only a single spin state (see Fig.~\ref{dos} (b)) which splits from the
valence band by 3.0 eV. Upon relaxation, it yields a distorted lattice in which the spatial distribution of the hole state is localized around a single molecule, as shown in Fig.~\ref{loc}. 

Next we analyze the ionization energies of the {\it localized} hole state ($E^+_{\rm loc}-E^0_{\rm ref}$) and the {\it delocalized} hole ($E^+_{\rm deloc}-E^0_{\rm ref}$), where $E^+$ and $E^0_{\rm ref}$ refer to the energy of the positively charged and the neutral ground state equilibrium system. Table~\ref{tab:energies} shows the results for benzene and anthracene crystal. The delocalized hole energy is computed in the neutral equilibrium geometry. After adding finite size corrections~\cite{corr}, we find $I_{\rm loc} \sim 7.85$ eV for benzene and 5.8 eV for anthracene, in good agreement with the experimental 7.58 and 5.7 eV associated with the hole polaron state~\cite{Sato, Silinsh}.  The localized hole states therefore exhibits a lower ionization potential by $E^{\rm st}=I_{\rm loc}-I_{\rm deloc}$ of $-0.67$ eV for benzene and $-1.07$ eV for anthracene.  This ``self-trapping energy'' exceed systematic uncertainties due to, say, electrostatic corrections~\cite{Makov} and clearly favors hole localization. We thus obtain a general prediction of {\it intrinsic self-trapping} of holes in a small polyacene crystals in the absence of impurities and crystal defects.

\begin{table}
\caption{Calculated ionization energy (in eV) of the localized and delocalized hole state and self-trapping energy of the polyacene crystals ($E^0_{\rm ref}$ is set zero). The absolute values of the ionization energy $I_{\rm loc}$ and $I_{\rm deloc}$ are obtained by adding a quadruple correction $\phi_q$ which is independent of the extent of charge delocalization. Experimental value is taken from Ref.~\cite{Sato} }
\begin{tabular}{c|cccccc|c}
polyacenes  &  $E^+_{\rm loc} $ &  $E^+_{\rm  deloc}$ & $\phi_q$ & $I_{\rm loc}$ &$I_{\rm deloc}$ &$E^{\rm st}$ &$I_{\rm Expt}$ \\\hline
benzene & 2.53 & 3.2 & 5.32 & 7.85 & 8.52& -0.67 &7.58 \\\hline
anthracene& -0.33 & 0.74 & 6.13  & 5.80  & 6.87& -1.07 & 5.70\\
\end{tabular}
\label{tab:energies} 
\end{table}
Compared to the gas phase, the stabilization energy for the delocalized hole in solid $\Delta I_{\rm deloc} = I_g- I_{\rm deloc}$ of 0.61 eV and 0.36 eV for benzene and anthracene corresponds to the energy associated with spreading out the charge between adjacent molecules and is consistent with the calculated bandwidth $\sim$ 0.5 eV of benzene and $\sim$ 0.34 eV of anthracene crystal~\cite{Hummer}. In contrast, the stabilization energy for the localized hole $\Delta I_{\rm loc}$ of 1.28 eV and 1.43 eV for benzene and anthracene compares well with the total polarization energy measured experimentally~\cite{Silinsh, Sato} and can be attributed to electronic polarization and geometry relaxation. The electronic polarization energy can be crudely estimated from a dielectric continuum model $\Delta P = -(1-1/\epsilon)(e^2/2R)$ where $R$ is the radius of the sphere surrounding the excess charge and $\epsilon$ is the dielectric constant~\cite{Silinsh}. 
For benzene we estimate $R$ on the order of  3-4 \AA~  (enclosing $>$ 95\% of the localized spin) from the spin distribution of the hole state in Fig.~\ref{loc} which yields a polarization energy of $\sim1-1.4$ eV in a range reported for polyacene crystals~\cite{Silinsh,Norton}. 
The geometry relaxation leads to an energy gain associated with the intramolecular relaxation~\cite{Marcus} of 0.17 eV for benzene and 0.14 eV for anthracene~\cite{note_rel}. The latter value is slightly higher than previous obtained from B3LYP hybrid functional~\cite{Norton,Coropceanu} but compares well with that from the linear coupling constant method~\cite{Martinelli}. There is also a cost of energy due to elastic lattice distortion (i.e., $E^0_{\rm loc}- E^0_{\rm ref}$) of $0.36$ eV for benzene and 0.15 eV for anthracene.  The self-trapping we observed is thus a result of competition between energy gain and cost associated with localization and delocalization. The values of these energy components are similar to those identified for molecular polarons~\cite{Silinsh} where the intramolecular relaxation energy corresponds to the formation energy and thus supporting the notion that the localized hole state is indeed a self-trapped hole polaron.  Predicting the outcome is not straightforward, especially in the nonlinear near-critical regime since the different energy components are not necessarily additive. The delicate balance highlights the importance of accurate DFT approaches in which the various energy contributions are properly accounted for.  

To investigate why the standard PBE and PBE0 functional fail to predict hole localization, we have also calculated the stabilization energy of the delocalized hole using the PBE and PBE0 functional. We find $\Delta I_{\rm deloc} = 3.34$ and 2.37 for $\alpha=0$ and 0.25, respectively, with a surprisingly strong dependence on $\alpha$. (The IP values obtained for (C$_6$H$_6$)$_n$ clusters, with $n$ up to 12, are given in the Supporting Information and confirm the periodic boundary condition predictions.)   Compared to 0.61 eV calculated at $\alpha = 0.7$, the values of $\Delta I_{\rm deloc}$ associated with $\alpha=0$ and 0.25 are much higher than the experimental total polarization energy in molecular crystals~\cite{Silinsh, Sato} and make the delocalized state unphysically favorable. This dovetails with our monomer-based analysis that $\alpha=0.7$ is needed to eliminate the many-electron SIE --- even though a completely delocalized hole exhibits a vanishing one-electron SIE.  To our knowledge, such strong IP dependence on DFT functionals has not been reported before. 
The theoretical reasoning for the shift observed in the delocalized hole
energy is related to the significant shift of the one-particle self energy
for the valence band maximum in solids when DFT is replaced with the GW
approximation~\cite{Hybertsen}. 

We have examined other small organic molecules besides polyacenes, including $p$-benzoquinone  (C$_6$H$_4$O$_2$) and 2,2'-bithiophene (C$_8$H$_6$S$_2$).  The critical value of $\alpha$ is somewhat system dependent, e.g., it is 0.6 for bithiophene and
0.5 for benzoquinone, all much greater than
$\alpha= 0.25$ used in the global PBE0 functional. The dependence of $\alpha_c$ on the
type of molecule seems to suggest specific differences in the response
of local environment, including the chemical species, molecular geometries, and bondings, upon ionization.  We have identified similar self-trapped hole polaron on a single molecular site in the associated crystals indicating a general 
behavior of excess charge in small molecular solids. Finally we note that transport mechanisms in molecular crystals, in particular, high mobility organic crystals, remain under debate~\cite{Troisi}. The ability of our approach to 
accurately describe the presence of polaron state in molecular crystals will be crucial for future studies of polaron dynamics purely from first principles. 

In summary, after showing that
DFT delocalization error is detrimental for the study of excess charges in organic molecular 
system, we suggest that this problem can be overcome with a hybrid density 
functional theory that incorporates an appropriate fraction of exact 
exchange. Our scheme enables an accurate description of the ionization 
and dimer dissociation limit of small organic molecules and predicts self-trapped hole polaron in a perfect molecular crystal that is not captured in the standard DFT. 
Our result suggests that semi-local and even standard hybrid DFT functionals overestimate the tendency of hole delocalization owing to errors in the delocalized state. 
The gas phase ionization potentials of small molecules, arguably related to
the energies of localized holes in solids, are however accurately predicted
by even the semi-local functionals that we have considered. Our study may contribute to the
understanding of possible mechanisms of deep-hole trap states~\cite{Bolinger} commonly
observed in organic electronics, and may potentially be
applicable to covalently bonded solids~\cite{Grossman, Lany, Wu,
Ramo, Franchini}. 

We thank P. Schultz, A. Wright, N. Modine, P. Feibelman, and X.Y. Zhu for helpful suggestions and comments.
This work is supported by the Energy Frontier Research Center funded by the U.S. DOE Office of Basic Energy Sciences under Award number DE-SC0001091. KL is also supported by the DOE under Contract DE-AC04-94AL85000. 
PFB was funded in part by the Welch Foundation (Grant \# F0020). Computational resources has been provided by New Mexico Computing Applications Center and Texas Advanced Computing Center.


\begin{thebibliography}{10}
\bibitem{corr}The localized hole energy is corrected for finite size effect by adding a monopole correction~\cite{mono}. For the delocalized hole, extrapolation to infinite cell size is attained by adopting fractional positive charges in a unit cell and studying the scaling behavior.
\bibitem{mono} The monopole correction is 0.55 eV and 0.42 eV with a dielectric constant $\epsilon=2.3$ and 3 for benzene and anthracene, respectively. A modification of the VASP zero potential~\cite{Leung} gives an additional respective $-1.2$ eV and $-1.35$ eV correction.
\bibitem{note_rel} The intramolecular relaxation energy depends somewhat on the fraction of the exact exchange, but the difference is much smaller than the self-trapping energy.
\end{thebibliography}
\end{document}